\documentstyle[11pt,paspconf,epsf]{article}

\def\eg{{\frenchspacing e.{\thinspace}g. }}

\def\et{{\it et~al.} }

\def\simlt{\hbox{ \rlap{\raise 0.425ex\hbox{$<$}}\lower 0.65ex\hbox{$\sim$} }}
\def\ltorder{\hbox{ \rlap{\raise 0.425ex\hbox{$<$}}\lower 0.65ex\hbox{$\sim$} }}
\def\simgt{\hbox{ \rlap{\raise 0.425ex\hbox{$>$}}\lower 0.65ex\hbox{$\sim$} }}
\def\gtorder{\hbox{ \rlap{\raise 0.425ex\hbox{$>$}}\lower 0.65ex\hbox{$\sim$} }}
\def\sles{\lower2pt\hbox{$\buildrel {\scriptstyle <}
   \over {\scriptstyle\sim}$}}
\def\sgreat{\lower2pt\hbox{$\buildrel {\scriptstyle >}
   \over {\scriptstyle\sim}$}}

\def\qquad{\quad\quad}

\newcommand{\be}{\begin{eqnarray}}
\newcommand{\ee}{\end{eqnarray}}

\begin{document}

\title{The GRAPE-4, a Teraflops Stellar Dynamics Computer}

\author{Piet Hut}
\affil{Institute for Advanced Study, Princeton, NJ 08540, U.S.A.}

\begin{abstract}
Recently, special-purpose computers have surpassed general-purpose
computers in the speed with which large-scale stellar dynamics
simulations can be performed.  Speeds up to a Teraflops are now
available, for simulations in a variety of fields, such as planetary
formation, star cluster dynamics, galactic nuclei, galaxy
interactions, galaxy formation, large scale structure, and
gravitational lensing.  Future speed increases for special-purpose
computers will be even more dramatic: a Petaflops version, tentatively
named the GRAPE-6, could be built within a few years, whereas
general-purpose computers are expected to reach this speed somewhere
in the 2010-2015 time frame.  Boards with a handful of chips from such
a machine could be made available to individual astronomers.  Such a
board, attached to a fast workstation, will then deliver Teraflops
speeds on a desktop, around the year 2000.
\end{abstract}

\section{Introduction}

Computational physics has emerged as a third branch of physics,
grafted onto the traditional pair of theoretical and experimental
physics.  At first, computer use seemed to be a straightforward
off-shoot of theoretical physics, providing solutions to sets of
differential equations too complicated to solve by hand.  But soon the
enormous quantitative improvement in speed yielded a qualitative shift
in the nature of these computations.  Rather than asking particular
detailed questions about a model system, we now use computers more
often to model the whole system directly.  Answers to relevant
questions are then extracted only after a full simulation has been
completed.  The data analysis following such a virtual lab experiment
is carried out by the computational physicist in much the same way as
it would be done by an experimenter of observer analyzing data from a
real experiment or observation.

With this shift from theory to experimentation, computers have become
important laboratory tools in all branches of science.  There is one
striking difference, though, between the use of a computer and that of
other types of lab equipment.  Whereas laboratory tools are typically
designed for a particular purpose, computers are usually bought off
the shelf, and used as is, without any attempt to customize them to
the particular usage at hand.  In contrast, it would be unthinkable
for a astronomy consortium to build a new observatory around a huge
pair of binoculars, as a simple scaled-up version of commercial
bird-watching equipment.

The reason for this difference in buying pattern has nothing to do
with an inherent difference between the activities of computing,
experimenting, or observing.  Building a special-purpose computer is
not more difficult than building a telescope, or any other major type
of customized laboratory equipment.  Rather, the difference in
attitude has everything to do with the fact that our computational
ability has gone through an extraordinary period of sustained rapid
exponential growth in speed.

Imagine that binoculars would grow twice as powerful every one or two
years.  If that were the case, astronomers might as well simply buy
the latest model binoculars, and use those for their observations.
Planning to build a big telescope would be self-defeating: in the ten
or so years it would take to design and build the thing, technology
would have progressed so much that commercial binoculars would
out-perform the special-purpose telescope.

Over the last forty years, computer speed has exponentially increased.
As a result, there has never been a particularly great need for
physicists to design and build their own computer.  As with all cases
of exponential growth, this tendency will necessarily flatten off.
How and when this flattening will occur is difficult to predict.  This
will depend on technological and economic factors that are as yet
uncertain.  But it is already the case that increase in computer speed
is significantly more modest than what could be expected purely from
the ongoing miniaturization of computer chips.  This trend, in the
case of general purpose computers, will be discussed briefly in
\S~\ref{GPC} \ Various alternatives, in the form of special-purpose
computing equipment, are mentioned in \S~\ref{SPC} \ One such
alternative, the GRAPE family of special-purpose computer hardware, is
reviewed in \S~\ref{GR} \ Some astrophysical applications of these
GRAPE machines are discussed in \S~\ref{AA} \ A preview of coming
GRAPE attractions is presented in \S~\ref{FUT}

\section{General-Purpose Computers}
\label{GPC}

After mainframes and minicomputers turned out to be no longer
cost-effective, some time around the early-to-mid eighties, the only
general-purpose computers used in physics were workstations and
supercomputers.  At first, there was an enormous gap in performance
between the two types of machines, but over the last fifteen years
this gap has narrowed steadily.

For example, during the eighties, supercomputers increased in speed by
about a factor of $10^2$, while microprocessors saw an increase of a
factor of $10^3$.  The main reason was that workstations at first were
rather inefficient, requiring many machine cycles for
a single floating point operation.  With increased chip size, this
situation improved rapidly.  In contrast, the first supercomputers,
built in the mid seventies, were designed specifically to deliver at
least one new floating point result for each clock cycle, through the
use of pipelines.

Although the speed of the floating point components for supercomputers
has continued to increase over the years, most of the increase in
their peak speed has been realized through increasing the number of
processing units.  This increase in parallelism has made the sharing
of memory by different processors increasingly cumbersome, involving
significant hardware overhead: a full interconnect between $N$
processors and a central memory bank requires an amount of additional
hardware that scales as $N^2$.

In contrast, the much faster speed-up of microprocessor-based
workstations has been possible exactly because there was (as yet) no
need for parallelism.  Throughout the eighties, chips did not
contain enough transistors to allow floating point operations to be
performed on a single chip in one cycle.  Therefore, personal
computers used to have a special floating point accelerator chip, in
addition to the central processor chip, and even this accelerator
typically needed several cycles even for the simplest operations of
addition and multiplication.  As a result, increase in the number of
transistors per chip translated linearly into an increase in speed.

However, this situation changed as soon as it became possible to put a
complete computer on a single chip, including a floating point unit
with the capability of producing a new output every cycle.  While in
itself a great achievement, this capability also creates new trouble.
From this point on, the scaling of general-purpose computers, based on
microprocessors, will become less favorable, for the following reasons.

With further miniaturization, a single chip will soon contain several
floating point units, with an extremely fast on-chip interconnect.
These interconnections, however, require a significant `real estate'
overhead on the chip: many extra components have to be added to the
chip in order to implement the administrative side of this fast
communication efficiently.  In addition, the off-chip communication
with the main memory is far slower, and tends to form a bottleneck.

As a result of both factors, a shrinking in feature size by a factor
two no longer guarantees a speed-up of a factor $\sim 8$, but rather
$2 \sim 4$.  In the eighties, when the feature width would become a
factor two smaller, four times as many transistors would fit on one
chip, and in addition the shrinking of the size of the transistors by
a factor two would allow a clock speed nearly twice as high as before.
However, this gain of a factor of eight from now on will be offset by
a communication penalty of a factor $2 \sim 4$.  The conclusion is
that microprocessors are now facing the same problem of increasing
`internal administrative bureaucracy' that supercomputer processors
have had to deal with for the last twenty years.

\section{Special-Purpose Computing Equipment}
\label{SPC}

\subsection{Special-Purpose Computers}

Until the late seventies, almost all scientific calculations were
carried out on general-purpose computers.  Around that time,
microprocessors began to offer a better price-performance ratio than
supercomputers.  By itself, this was not very helpful to a physicist,
given the fact that a single microprocessor could only offer a speed
of 10 kflops or so, peanuts compared to the supercomputers of those
days, with peak speeds above 100 Mflops.  The key to success was to
find a way to combine the speed of a large number of those cheap
microprocessors.

This was exactly what several groups of physicists did, in the
eighties.  They took large numbers of off-the-shelf microprocessors,
and hooked them up together.  Building these machines was not too
hard, and indeed raw speeds at low prices were reached relatively
easily.  The main problem was that of software development.  To get a
special-purpose machine to do a relevant physics calculation, and to
report the results in understandable form, provided formidable
challenges.  For example, writing a reasonably efficient compiler for
such a machine was a tedious and error-prone job.  In addition,
developing application programs was no simple task either.

An interesting and somewhat unexpected development has been the
commercialization of these machines, originally built by and for
physicists.  The design of most of the current highly-parallel
general-purpose computers has been directly or indirectly influenced
by the early special-purpose computers.  This blurring of the
distinction between special-purpose and general-purpose computers may
continue in the future, when demand for higher peak speeds will force
increasing parallelization to occur.

This development reflects the fact that the so-called special-purpose
machines in physics actually attacked a general type of problem: how
to let many individual processors cooperate on a single computational
task.  The fact that the applications have been rather specialized in
many cases (to particle physics, astrophysics, or hydrodynamics) is
less important than the fact that each application required a
carefully balanced strategy at dynamic inter-processor communication.
As a result, the experience gained from the development of both
hardware and software for special-purpose computers has turned out to
be very helpful for the development of their general-purpose
counterparts as well.

\subsection{Special-purpose Accelerators}

In the late eighties, an alternative model was developed.  Following
the example of some special-purpose components designed as back-end
processors in radio telescopes, the idea was advanced to design
special hardware components to speed up critical stages within
large-scale simulations, most of which would still be delegated to
general-purpose workstations.

A similar idea had already been employed for general-purpose computers
as well.  In the early eighties, personal computers would come with a
central processor that could handle floating-point calculations only
in software, at rather low efficiency.  Significant speed-up, of an
order of magnitude, could be obtained by including a so-called
floating-point accelerator, at only a fraction of the cost of the
original computer.  Another example is the use of graphics
accelerators in most modern personal computers.

Building a special hardware accelerator for a critical segment of a
physics simulation is another example of this general approach.  In
this way, the good cost-performance ratio of special-purpose hardware
can be combined with the flexibility of existing workstations, without
much of a need for special software development.  This approach can be
compared to using hand-coded assembly-language or machine-code for an
inner loop in an algorithm that otherwise is programmed in a
higher-level language --- the difference being that this inner loop is
now realized directly in silicon.

\section{The GRAPE project}
\label{GR}

\subsection{Prehistory}

In 1984 a group of astrophysicists and computer scientists built the
digital Orrery, a 10 Mflops special-purpose computer designed to
follow the long-term evolution of the orbits of the planets (Applegate
\et 1985).  For that purpose, ten processors were connected in a ring,
one for each planet (or test particle).  The processors were designed
around an experimental 64-bit floating-point chip set developed by
HP. Each chip could perform one floating point operation in 1.25
$\mu$s. A central controller send instructions to all processors at
each machine cycle.

A few years later, results from the Orrery lead to the important
discovery of the existence of a positive Lyapunov coefficient for the
evolution of the orbit of Pluto, which was interpreted as a sign of
chaos (Sussman \& Wisdom 1988).

Besides the question of the long-term evolution of planetary orbits,
there were many other problems in gravitational dynamics that required
far more than the typical speed available to astrophysicists in the
mid-eighties.  While significant speed-up was obtained with the
introduction of more efficient algorithms (\eg\ Barnes \& Hut 1986,
1989), many problems in stellar dynamics could not be effectively
tackled with the hardware available at that time.

Among those problems, the most compute-intensive was the long-term
simulation of star clusters past core collapse.  The record
in that area in the late eighties was held by Makino \& Sugimoto
(1987) and Makino (1989), for $N$-body calculations with $N=1000$ and
$N=3000$, respectively.  Unfortunately, the computational costs for
these types of calculations scales roughly with $N^3$, which meant
that realistic simulations of globular clusters, with $N$ in the range
$10^5\sim10^6$, were still a long way off.

The only hope to make significant progress in this area was to make
use of the fastest supercomputers available, in the most efficient way
possible.  Therefore, the next step we took was a detailed analysis of
the algorithms available for the study of dense stellar systems
(Makino \& Hut 1988, 1990), following the earlier analysis given by
Makino (1986).

Our analysis showed that the best integration schemes available, in
the form of Aarseth's individual-timestep predictor-corrector codes
(Aarseth 1985), were close to the theoretical performance limit.
Based on these results, we predicted that a speed of order 1 Teraflops
would be required to model globular star clusters, and to verify the
occurrence of gravothermal oscillations in such models (Hut \et 1988).

Unfortunately, such speeds were not commercially available in those
days, and it was clear that they would not be available for another
ten years or so.  The fastest machine that we could lay our hands on
was the Connection Machine CM-2, which was first being shipped by
Thinking Machines in 1987.  In the Fall of that year, Jun Makino and I
spent a few months at Thinking Machines, to perform an in-depth
analysis of the efficiency of various algorithms for stellar dynamics
simulations on the CM-2.

The results were somewhat disappointing (Makino and Hut 1989), in that
most large-scale simulations could utilize only $\sim 1$\% of the
peak-speed of the CM-2.  As a result, even with a formidable peak
speed of tens of Gigaflops, most of our simulations only obtained a
speed of a few hundred Megaflops, when scaled up to a full CM-2
configuration.  The main reason for its poor performance was the
slowness of the communication speed compared to the speed of the
floating point calculations.

Since we needed a Teraflops in order to study gravothermal
oscillations and other phenomena in dense stellar systems, it was
rather disheartening that we could not even reach an effective
Gigaflops.  And given the typical increase in speed of supercomputers,
by a factor of $\sim 10$ every five years, it seemed clear that we
would have to wait till well after the year 2000, before being able to
compute at an effective Teraflops speed.

In reaction to our experiences, Sugimoto took up the challenge and
formed a small team at Tokyo University to explore the feasibility of
building special-purpose hardware for stellar dynamics simulations.
This group started their project in the Spring of 1989, resulting in
the completion of their first machine in the Fall of that same year
(Ito \et 1990).

\subsection{The GRAPE Family}

The name GRAPE stands for GRAvity PipE, and indicates a family of
pipeline processors that contain chips specially designed to calculate
the Newtonian gravitational force between particles.  A GRAPE
processor operates in cooperation with a general-purpose host
computer, typically a normal workstation.  The force integration and
particle pushing are all done on the host computer, and only the
inter-particle force calculations are done on the GRAPE.  Since the
latter require a computer processing power that scales with $N^2$,
while the former only require $\propto N$ computer power, load balance
can always be achieved by choosing $N$ values large enough.

The development history of the Grape series of special-purpose
architectures shows a record of rapid performance improvements (see
Table 1).  The limited-precision Grape-1 achieved 240 Mflops in 1989;
its successor, the Grape-3, reached 15 Gflops in 1991.  Over 30
Grape-3 systems are currently in use worldwide in applications (such
as tree codes and SPH applications) where high numerical precision is
not a critical factor.

A prototype board of the full-precision Grape-2 achieved 40 Mflops in
1990.  The full Grape-4 system reached 1.1 Teraflops (peak) in 1995.
Individual Grape-4 boards, delivering from 3 to 30 Gflops depending on
configuration, are currently in use at 5 institutions around the
world.

A third development track is represented by the GRAPE-2A and MD-GRAPE
machines, which include a user-loadable force look-up table that can
be used for arbitrary central force laws (targeted at molecular
dynamics applications).  Overall, the pace of development has been
impressive: 10 special-purpose machines with a broadening range of
applications and a factor of 4000 speed increase in just over 6 years.

The Grape-4 developers have won the Gordon Bell prize for
high-performan\-ce computing in each of the past two years.  In 1995,
the prize was awarded to Junichiro Makino and Makoto Taiji for a sustained
speed of 112 Gflops, achieved using one-sixth of the full machine on a
128k particle simulation of the evolution of a double black-hole
system in the core of a galaxy.  The 1996 prize was awarded to
Toshiyuki Fukushige and Junichiro Makino for a 332 Gflops simulation of the
formation of a cold dark matter halo around a galaxy, modeled using
768k particles on three-quarters of the full machine.

\vbox{
\centerline{Table 1}
\smallskip
\centerline{Summary of GRAPE Hardware}
\bigskip
\vbox{\hbox to \hsize{\hfil\vbox{\tabskip=1 em \halign
  {#\hfil&\hfil#\hfil&\hfil#\hfil&#\hfil\cr
\noalign{\hrule}\cr
\noalign{\smallskip}
\noalign{\hrule}\cr
\noalign{\medskip}
\quad{\it Limited-Precision Data Path}\hidewidth & & & \cr
\noalign{\medskip}
Machine  & Year & Peak Speed & Notes\cr
\noalign{\smallskip}
GRAPE-1  & 1989 & 240 Mflops       & Concept system, GPIB interface\cr
GRAPE-1A & 1990 & 240 Mflops       & VME interface\cr
GRAPE-3  & 1991 & 15 Gflops        & 48 Custom LSIs, 10 MHz clock\cr
GRAPE-3A & 1993 & 5 Gflops/board   & 8 chip version for distribution,\cr
         &      &                  & 20 MHz, PCB implementation\cr
\noalign{\bigskip}
\noalign{\medskip}
\quad{\it Full-Precision Data Path}\hidewidth & & & \cr
\noalign{\medskip}
Machine  & Year & Peak Speed & Notes\cr
\noalign{\smallskip}
GRAPE-2  & 1990 & 40 Mflops        & IEEE precision, commercial\cr
         &      &                  & chips\cr
HARP-1   & 1993 & 180 Mflops       & ``Hermite'' pipeline\cr
HARP-2   & 1993 & 2 Gflops         & Evaluation system of the\cr
         &      &                  & custom chips to be used in\cr
         &      &                  & GRAPE-4\cr
GRAPE-4  & 1995 & 1.1 Tflops       & The Teraflops GRAPE, 1692\cr
         &      &                  & pipelines\cr
\noalign{\bigskip}
\noalign{\medskip}
\quad{\it Arbitrary Force Law}\hidewidth & & & \cr
\noalign{\medskip}
Machine  & Year & Peak Speed & Notes\cr
\noalign{\smallskip}
GRAPE-2A & 1992 & 180 Mflops       & Force look-up table\cr
MD-GRAPE & 1995 & 4 Gflops         & Custom chip with force look-up\cr
         &      &                  & table\cr
\noalign{\medskip}
\noalign{\hrule}\cr
\noalign{\medskip}}}\hfil}}
}

\bigskip

\subsection{Using the GRAPE}

Modifying an existing program to use the GRAPE hardware is
straightforward, and entails minimal changes.  Subroutine and function
calls (written in C or FORTRAN) to the GRAPE hardware replace the
force-evaluation functions already found in existing $N$-body codes.

Communication between host and GRAPE is accomplished through a
collection of about a dozen interface routines.  The force evaluation
code which is replaced typically consists of only a few dozen lines at
the lowest level of an algorithm.  Thus, using the GRAPE calls only
for small, localized changes which in no way inhibit future
large-scale algorithm development.

The GRAPE interface has been successfully incorporated into the
Barnes-Hut tree algorithm (Barnes \& Hut 1986; Makino 1991) and the
P$^3$M scheme (Hockney \& Eastwood 1988; Brieu, Summers, \& Ostriker
1995).

\vfill\eject

Here is a typical code fragment for the Newtonian force calculation on
a workstation:

\bigskip

{
\obeylines
\tt
\medskip
\ \ \ \ \ \ subroutine\ accel\_workstation
\ \ \ \ \ \ do\ 10\ k\ =1,ndim	
\ \ \ \ \ \ \ \ do\ 20\ i=1,nbody
\ \ \ \ \ \ \ \ \ \ accnew(i,k)=0.0
\ 20\ \ \ \ \ continue
\ 10\ \ \ continue
\ \ \ \ \ \ do\ 30\ i=1,nbody-1	
\ \ \ \ \ \ \ \ do\ 40\ j=i+1,nbody
\ \ \ \ \ \ \ \ \ \ do\ 50\ k\ =\ 1,3
\ \ \ \ \ \ \ \ \ \ \ \ dx(k)=pos(k,j)-pos(k,i)
\ 50\ \ \ \ \ \ \ continue
\ \ \ \ \ \ \ \ \ \ r2inv=1.0/(dx(1)**2+dx(2)**2+dx(3)**2+eps2)
\ \ \ \ \ \ \ \ \ \ r3inv=r2inv*sqrt(r2inv)
\ \ \ \ \ \ \ \ \ \ do\ 60\ k=1,3
\ \ \ \ \ \ \ \ \ \ \ \ accnew(k,i)=accnew(k,i)+r3inv*mass(j)*dx(k)
\ \ \ \ \ \ \ \ \ \ \ \ accnew(k,j)=accnew(k,j)-r3inv*mass(i)*dx(k)
\ 60\ \ \ \ \ \ \ continue
\ 40\ \ \ \ \ continue
\ 30\ \ \ continue
\ \ \ \ \ \ end
}

\bigskip\bigskip
\noindent
To use the grape, all that has to be done is to replace the inner loop
of the force calculations by a few special function calls in order to
offload the bulk of the computation onto the GRAPE hardware:

\bigskip

{
\obeylines
\tt
\medskip
\ \ \ \ \ \ subroutine\ accel\_grape
\ \ \ \ \ \ call\ g3init()
\ \ \ \ \ \ xscale\ =\ 1.0d0/1024
\ \ \ \ \ \ call\ g3setscales(xscale,\ mass(1))
\ \ \ \ \ \ call\ g3seteps2(eps2)
\ \ \ \ \ \ call\ g3setn(nbody)
\ \ \ \ \ \ \ \ do\ 20\ i=1,nbody
\ \ \ \ \ \ \ \ \ \ \ call\ g3setxj(i-1,pos(1,i))
\ \ \ \ \ \ \ \ \ \ \ call\ g3setmj(i-1,mass(i))
\ 20\ \ \ \ \ \ continue
\ \ \ \ \ \ nchips=g3nchips()
\ \ \ \ \ \ do\ 30\ i=1,nbody,nchips
\ \ \ \ \ \ \ \ \ ii\ =\ min(nchips,\ nbody\ -\ i\ +\ 1)
\ \ \ \ \ \ \ \ \ call\ g3frc(pos(1,i),accnew(1,i),pot(i),ii)
\ 30\ \ \ continue
\ \ \ \ \ \ call\ g3free
\ \ \ \ \ \ end
}

\section{Some Astrophysical Applications}
\label{AA}

In this brief review, there is no room for an exhaustive review of the
scientific results that have been obtained with the few dozen GRAPE
machines that have been installed in a number of different research
institutes around the world.  In addition to the four fields listed
below, the GRAPEs have been used in a variety of other areas, for
example to study the role of exponential divergence of neighboring
light trajectories on gravitational lensing, the formation of
large-scale structure in the Universe, the role of violent
relaxation in galaxy formation, and the effectiveness of hierarchical
merging in galaxy clusters.

\subsection{Planet Formation.}

Ida \& Makino (1992a,b) used the GRAPE-2 to investigate the evolution
of the velocity distribution of a swarm of planetesimals, with an
embedded protoplanet.  They confirmed that equipartition is achieved
and that therefore runaway growth should take place, along the lines
suggested by Stewart \& Wetherill (1988).

Kokubo \& Ida (1995) used the HARP-2 (a smaller prototype of the
GRAPE-4) to simulate a system of two protoplanets and many
planetesimals.  They found that the separation between two planets
tends to grow to roughly 5 $r_H$ (the Hill radius). They coined the
term `orbital repulsion' for this phenomenon, and provided a
qualitative explanation for its occurrence.

Kokubo \& Ida (1996a) used the GRAPE-4 to simulate planetary growth
assuming perfect accretion, where any physical collision leads to
coalescence.  They started with 3000 equal-mass planetesimals.  After
20,000 orbits, they found that the most massive particle had become
300 times heavier, while the average mass of the particles increased
by only a factor of two.  Kokubo \& Ida (1996b) extended these
calculations. They showed that several protoplanets are formed and grow
while keeping their mutual separations within the range 5-10
$r_H$. Their results strongly suggests that orbital repulsion has
determined the present separation between the outer planets.

\subsection{Star Cluster Evolution.}

The first scientific result obtained with the GRAPE-4 was the
demonstration of the existence of gravothermal oscillations in
$N$-body simulations.  Predicted more than ten years earlier by
Sugimoto \& Bettwieser (1983), they were found by Makino (1996a), and
presented by him at the I.A.U. Symposium 174 in Tokyo, in August 1995
(Makino 1996b).  Using more than 32,000 particles, he was also able to
confirm the semi-analytical predictions made by Goodman (1987).  The
calculation took about two months, using only one quarter of a full
GRAPE-4, running at a speed of 50 Gflops.

We are currently exploring ways to couple stellar dynamics and stellar
evolution in one code, in order to perform more realistic simulations
of star cluster evolution.  Based on steller evolution recipes
implemented by Portegies Zwart \& Verbunt (1996), we have carried out
a series of increasingly realistic approximations (Portegies Zwart \et
1997a,b,c); see our web site with a movie that shows a star cluster,
as an evolving $N$-body system side-to-side with its correspondingly
evolving H-R diagram, at http://casc.physics.drexel.edu

\subsection{Density Profiles of Galactic Nuclei.}

Ebisuzaki \et (1991) used the GRAPE-2 to simulate the merging of two
galaxies, each with a central black hole, using up to $4096+2$
particles.  They found an increase in core radius, as a result of the
heating of the central regions caused by the spiral-in of the two black
holes.

Makino and Ebisuzaki (1996) used the GRAPE-4 to study hierarchical
merging, in which the merger product of one pair of galaxies was used
as a template for constructing progenitors for the next simulation of
merging galaxies.  They used more than 32,000 particles.  They found
the ratio between the core radius and the effective radius to converge
to a value depending on the mass of the black holes.

However, it turned out that 32k particles were not enough.  Makino
(1997) performed a similar type of calculation with 256k particles,
and found a core structure which was rather different from that
obtained in the previous 32k runs.  In particular, he found the volume
density of stars to decrease in the vicinity of the black hole binary
in the 256k runs, and ascribed this to the `loss cone' effect
predicted by Begelman \et (1980).

Fukushige \& Makino (1997) used the GRAPE-4 to simulate hierarchical
clustering, using an order of magnitude more particles than in
previous studies.  They found that the central density profiles are
always steeper than $\rho \sim r^{-1}$.  They interpreted the observed
shallower cusps as the result of the spiral-in of the central black
holes from the progenitor galaxies, involved in the merging process.

\subsection{Interactions between Galaxies.}

Okumura \et (1991) used the GRAPE-1 to investigate the structure of
merger remnants formed from encounters between two Plummer models on
parabolic orbits, using 16,000 particles.  They determined the
non-dimensional rotation velocity $V_{max}/\sigma_0$, where $V_{max}$
denotes the maximum rotation velocity and $\sigma_0$ is the velocity
dispersion at the center.  They found typical values of $\sim 0.6$ for
merging at large initial periastron separations.  Their result is in
good agreement with the observation of large ellipticals, which show a
rather sharp cutoff in the distribution of $V_{max}/\sigma_0$ around
0.6.

Makino \& Hut (1997) used the GRAPE-3A to simulate more than 500
galaxy encounters, in order to determine their merger rate as a
function of incoming velocity, for a variety of galaxy models.  They
characterized the overall merger rate in a galaxy cluster by a single
number, derived from their cross sections by an integration over
encounter velocities in the limit of a constant density in velocity
space.  In addition, they provided detailed information concerning the
reduction of the overall encounter rate through tidal effects from the
cluster potential as well as from neighboring galaxies.

\section{Coming Attractions: the GRAPE-6}
\label{FUT}

In the GRAPE-4, once all pipelines are filled, each chip produces one
new inter-particle interaction (corresponding to $\sim60$
floating-point operations) every three clock cycles.  For a clock
speed of 30 MHz, a peak chip speed of $\sim0.6$ Gflops is achieved.
The GRAPE-4 chips represent 1992 technology (1 $\mu$m fabrication line
width).  Even if no changes were made in the basic design, advances in
fabrication technology would permit more transistors per chip and
increased clock speed, enabling a 50--100 MHz, 10--30 Gflops chip with
1996 (0.35 $\mu$m line width) technology, and a 100--200 MHz, 50--200
Gflops chip with 1998 (0.25 $\mu$m) technology.  Based on these
projected performance improvements, a total of $\sim10^4$ GRAPE-6
chips of 100 Gflops each could be combined to achieve Petaflops speeds
by the year 2000, for a total budget of 10 million dollars.  
We have recently completed an initial `point design study' of the
feasibility of constructing such a system (McMillan \et 1996).
This study was funded by the NSF, in conjunction with NASA and
DARPA, as part of a program aimed at paving the way towards
Petaflops computing.

While planning to build a hardwired Petaflops-class computational
engine, we are also investigating complementary avenues, based on the
use of reconfigurable logic, in the form of Field-Programmable Gate
Array (FPGA) chips.  The merging of custom LSI and reconfigurable
logic will result in a unique capability in performance and
generality, combining the extremely high throughput of special-purpose
devices with the flexibility of reconfigurable systems.  In
many applications, gravity requires less than 99\% of the computing
power.  Although the remainder of the CPU time is typically dominated
by just one secondary bottleneck, its nature varies greatly from
problem to problem.  It is not cost-effective to attempt to design
custom chips for each new problem that arises.  In these
circumstances, a FPGA-based system can restore the balance, and
guarantee scalability from the Teraflops to the Petaflops domain,
while still retaining significant flexibility.  Astrophysical
applications could include, for example, various forms of Smooth
Particle Hydrodynamics (SPH), for applications ranging from
colliding stars to the formation of large-scale structure in the
Universe.

An additional benefit of the construction of Petaflops-class machines
will be the availability of individual chips at reasonable prices,
once the main machine has been designed and constructed.  A typical
GRAPE-6 chip will run at $\sim 100$ Gflops.  A single board with 10
or more chips will already deliver a speed of 1 Teraflops or more, for
a total price that is likely to lie in the range of 10,000 -- 20,000
dollars.  Hooking such a board up to a workstation will instantly
change it into a top-of-the-line supercomputer.

\acknowledgments
I thank Jun Makino and Steve McMillan for their comments on the manuscript.
This work was supported in part by the National Science Foundation
under grant ASC-9612029.

\end{document}